\documentclass[traditabstract]{aa} 
\usepackage{graphicx}
\usepackage{txfonts}
\usepackage{natbib}

\authorrunning{Brunthaler et al.}

\newcommand{\kms}{km~s$^{-1}$}
\newcommand{\nasyr}{$\mu$as~yr$^{-1}$}

\begin{document}
   \title{VLBI observations of SN\,2008iz:}

   \subtitle{I. Expansion velocity and limits on anisotropic expansion}

   \author{A. Brunthaler\inst{1}
          \and
	  I. Mart\'i-Vidal\inst{1}
	  \and
          K.M. Menten\inst{1}
          \and 
          M.J. Reid\inst{2}
          \and 
          C. Henkel\inst{1}
          \and 
          G.C. Bower\inst{3}
          \and 
          H. Falcke\inst{4,5}
          \and 
          H. Feng\inst{6}
          \and
          P. Kaaret\inst{7}
	  \and
          N.R. Butler\inst{3} 
          \and 
          A.N. Morgan\inst{3}
          \and
	  A. Wei{\ss}\inst{1}
          }

   \institute{Max-Planck-Institut f\"ur Radioastronomie, Auf dem H\"ugel 69,
              53121 Bonn, Germany\\
              \email{brunthal@mpifr-bonn.mpg.de}
         \and
             Harvard-Smithsonian Center for Astrophysics, 60 Garden Street,
              Cambridge, MA 02138, USA
         \and
              UC Berkeley, 601 Campbell Hall, Astronomy Department \& Radio
              Astronomy Lab, Berkeley, CA 94720, USA  
         \and
               Department of Astrophysics, Radboud Universiteit
               Nijmegen, Postbus 9010, 6500 GL Nijmegen, the Netherlands
               \and
              ASTRON, Postbus 2, 7990 AA Dwingeloo, the Netherlands
         \and
              Department of Engineering Physics and Center for Astrophysics, Tsinghua University, Beijing 100084, China
         \and
              Department of Physics and Astronomy, University of Iowa, Van Allen Hall, Iowa City, IA 52242, USA
             }

   \date{Received }

 
  \abstract
  {We present observations of the recently discovered supernova 2008iz in M82 
   with the VLBI High Sensitivity Array at 22 GHz, the Very Large Array at 
   frequencies of 1.4, 4.8, 8.4, 22 and 43 GHz, and the Chandra X-ray 
   observatory. The supernova was clearly detected on two VLBI images, 
   separated by 11 months. The source shows a ring-like morphology and 
   expands with a velocity of $\sim$ 23000 \kms. The most likely explosion date
   is in mid February 2008. The measured expansion speed is a factor of $\sim$2
   higher than expected under the assumption that synchrotron self-absorption 
   dominates the light curve at the peak, indicating that this absorption 
   mechanism  may not be important for the radio emission. We find no evidence 
   for an asymmetric explosion. The VLA spectrum shows a broken power law, 
   indicating that the source was still optically thick at 1.4 GHz in April 
   2009. Finally, we report upper limits on the X-ray emission from SN\,2008iz 
   and a second radio transient recently discovered by MERLIN observations.
}


   \maketitle
%

\section{Introduction}

Radio supernovae (RSNe) are rare events and difficult to study. So far only 
about two dozen have been detected \citep[e.g][]{Weiler2002} and most of them 
are quite distant and rather weak. Only few radio supernovae have been imaged 
with Very Long Baseline Interferometry (VLBI) techniques and for only four 
has it been possible to study the evolution of the expanding shell 
(SN\,1979C, SN\,1986J, SN\,1987A, SN\,1993J). The best studied radio supernova 
so far is SN\,1993J in M81. Following the expansion of the supernova allowed 
many different phenomena to be studied \citep{Marcaide1997, Marcaide2009, 
Bietenholz2001, Bietenholz2003, Perez2001, Perez2002, Bartel2002, Bartel2007}, 
including a measurement of the expansion speed, the deceleration of the shock 
front, and the proper motion of the supernova shell, for which a limit was 
obtained.

The recent discovery of a new bright radio supernova in M82, SN\,2008iz, 
\citep{BrunthalerMentenReid2009,BrunthalerMentenReid2009b} at a similar 
distance as SN\,1993J offers the rare opportunity to study the evolution of 
another supernova in great detail and to make a comparison to SN\,1993J. So 
far, SN\,2008iz was only detected in the radio band, with the
VLA at 22 GHz \citep{BrunthalerMentenReid2009}, MERLIN at 5 GHz 
\citep{BeswickMuxlowPedlar2009}, and the Urumqi telescope at 5 GHz 
\citep{MarchiliMartiVidalBrunthaler2009}. There are no reported detections in 
visible light and \cite{FraserSmarttCrockett2009} report only a non detection 
in the near infrared on 2009 June 11. The non-detections at other wavebands 
indicate that the supernova exploded behind a large gas or dust cloud in the
central part of M82. Thus, it has not been possible to classify this supernova.
However, since type Ia supernovae are not known to show strong radio emission, 
SN\,2008iz is most likely a core collapse supernova, i.e. type Ib/c or II.

Here we present the first VLBI images of SN\,2008iz taken $\sim$2.5 and 
$\sim$13.5 month after the explosion, a radio spectrum (at an age of 
$\sim$14.5 month) from 1.4 to 43 GHz and Chandra X-ray observations. 
Throughout the paper, we adopt a distance of 3.6 Mpc \citep[based on a Cepheid 
distance to M81 determined by][] {FreedmanHughesMadore1994}.


\section{Observations and data reduction}

 \subsection{High Sensitivity Array observations at 22 GHz}

M82 was observed using the High Sensitivity Array including the
NRAO\footnote{The National Radio Astronomy Observatory is a facility of the 
National Science Foundation operated under cooperative agreement by Associated 
Universities, Inc.} Very Long Baseline Array (VLBA), the Very Large Array 
(VLA), the Green Bank Telescope (GBT), and the Effelsberg 100m telescope under
project code BB255 on 2008 May 03 and 2009 April 08. The total observing time 
at each epoch was 12 hours. We used M81*, the nuclear radio source in M81, as 
phase calibrator and switched between M81*, M82, and 3 extragalactic background
quasars every 50 seconds in the cycle M81* -- 0945+6924 -- M81* -- 0948+6848 
-- M81* -- M82 -- 1004+6936 -- M81*, yielding an integration time of $\sim$100
minutes on M82. J1048+7143 was observed every $\sim$30 minutes to phase-up 
the VLA. Furthermore, DA193 was observed as fringe finder. The data were 
recorded with four 8 MHz frequency bands in dual circular polarization, with 
Nyquist sampling and 2 bits per sample (i.e., a total recording rate of 
256 Mbit~s$^{-1}$).

Before, in the middle, and at the end of the phase referencing observations,
we included {\it geodetic blocks}, where we observed 18--21 bright sources from
the International Celestial Reference Frame (ICRF) at 22 GHz for $\sim$75 
minutes to measure the tropospheric zenith delay offsets at each antenna 
\citep[for a detailed discussion see][]{ReidBrunthaler2004,
BrunthalerReidFalcke2005b}. The geodetic blocks were recorded with 8 IFs of 
8 MHz in left circular polarization. The first IF was centered at a frequency 
of 22.01049 GHz, and the other seven IFs were separated by 12.5, 37.5, 100.0, 
262.5, 312.5, 412.5, and 437.5 MHz, respectively. Here, only a single VLA 
antenna was used, which could observe only five IFs at 0.0, 12.5, 37.5, 412.5, 
and 437.5 MHz relative to 22.01049 GHz.

The data were correlated at the VLBA Array Operations Center in Socorro, New
Mexico. In the observation on 2008 May 03, the data  were correlated at the
position of known water maser emission 
(09$^\mathrm{h}$55$^\mathrm{m}$51$^\mathrm{s}$.38702,
+69$^\circ$40$'$44$''$.4676, J2000) with 128 spectral channels per IF and an
integration time of 1 second. The observation on 2009 April 08 was correlated
at the position 09$^\mathrm{h}$55$^\mathrm{m}$51$^\mathrm{s}$.5500, 
+69$^\circ$40$'$45$''$.792, (J2000),
close to SN\,2008iz  with 16 spectral channels and an integration time of 1
second. 

The data reduction was performed with the NRAO Astronomical Image Processing 
System 
(AIPS) and involved standard steps. First we shifted the position of M82 to 
the position of SN\,2008iz using CLCOR in the observation on 2008 May 03. We 
applied the latest values of the Earth's orientation parameters and performed 
zenith delay corrections based on the results of the geodetic-like 
observations. Total electron content maps of the ionosphere were used to 
correct for phase changes by the ionosphere. A-priori amplitude calibration 
was applied using system temperature measurements and standard gain curves. 
We performed a ``manual phase-calibration'' using the data from J1048+7143 to 
remove instrumental phase offsets among the frequency bands. Then, we 
fringe fitted the data from M81* and performed an amplitude 
self-calibration on M81*. In the observation on 2008 May 03, we also performed,
first, a phase-self calibration and later an amplitude self-calibration on
SN\,2008iz. In the observation on 2009 April 08, the inclusion of the phased 
VLA increased the noise in the images significantly, possibly due to a problem 
with the phasing of the array (the VLA was in B-configuration). Thus, we 
flagged the VLA data from this epoch.

\subsection{VLA observation on 2009 April 27 (1.4-43 GHz)}

We observed M82 with the VLA on 2009 April 27 at 1.4, 4.8, 8.4, 22, and 43 GHz.
The total observing time was 4 hours. We observed with two frequency bands of 
50 MHz, each in dual circular polarization. 3C\,48 and J1048+7143 were used as 
primary flux density and phase calibrators, respectively. At 1.4, 4.8, 
and 8.4 GHz, we used a switching cycle of six minutes, spending one minute on 
the phase calibrator and five minutes on M82. We repeated these cycles 5 times 
over the observations, yielding an integration time of $\sim$25 minutes at 
each frequency. At 22 and 43 GHz, we used a switching cycle of three minutes, 
spending one minute on the phase calibrator and two minutes on M82. These 
cycles were repeated 10 times during the observation, yielding an 
integration time of $\sim$20 minutes at both frequencies. 

The data reduction was performed in AIPS and involved amplitude calibration of 
3C\,48 using source models. Then we calibrated the phases using J1048+7143 and 
made one phase and amplitude self-calibration on J1048+7143. The calibration 
was then transferred to the target source M82. We performed one phase 
self-calibration on M82 at all frequencies except 43 GHz, where the source is 
too weak. 

\begin{figure}
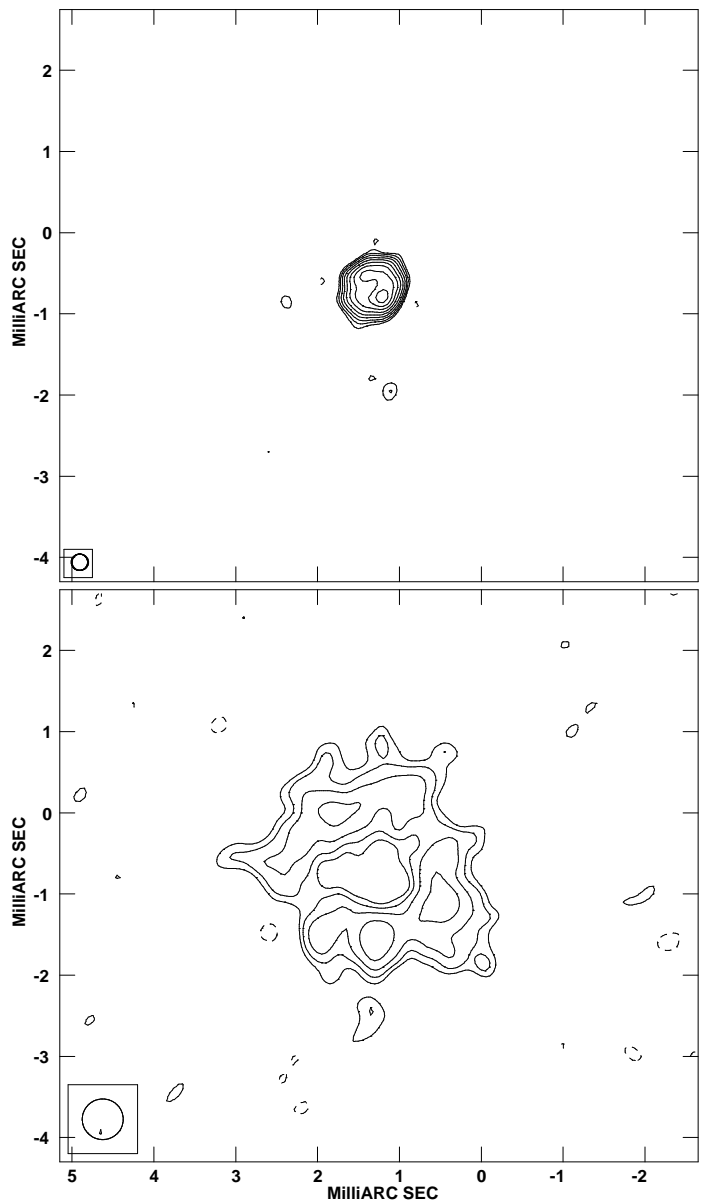

   \resizebox{0.5\textwidth}{!}
            {\includegraphics[bb=1.5cm 7.40cm 20.0cm
              22.8cm,clip,angle=0]{f1a.ps}}
   \resizebox{0.5\textwidth}{!}
            {\includegraphics[bb=1.5cm 6.50cm 20.0cm
              22.8cm,clip,angle=0]{f1b.ps}}
      \caption{HSA contour plots of SN\,2008iz at 22 GHz in May 2008 (top) and 
        April 2009 (bottom). Contours start at 0.3 mJy (top), 0.1 mJy 
        (bottom) and increase with $\sqrt2$. The images are restored with a 
        circular beam of FWHM 0.2 mas (top) and 0.5 mas (bottom). The original
        beam sizes were 0.31$\times$0.23 mas at a position angle of $-34^\circ$
        at the first epoch and 0.39$\times$0.27 at a position angle of 
        $-40^\circ$ at the second epoch.
   }
         \label{fig:hsa}
   \end{figure}

\subsection{Chandra and Swift XRT observations}

The Chandra X-ray observatory (CXO) observed M82 on 2008 October 4, 
and 2009 April 17 and 29. Each observation was taken with an exposure of about 
18 ks. In these observations, the target is off the optical axis by more than 
3 arcmin, where the point spread function looks like an extended ellipse 
covering multiple pixels.

M82 was also observed with the Swift X-ray Telescope 
\citep[XRT;][]{GehrelsChincariniGiommi2004,BurrowsHillNousek2005}
on 2007 January 26, 2008 May 1, and 2009
April 25 with exposures of 4.6 ks, 5.0 ks, and 4.7 ks, respectively.
Due to the low resolution of the telescope, the collection of known
X-ray sources looks point-like in the XRT data, so count rates from
the conglomerate of sources were measured.  The data suffer from heavy
pile-up, which is accounted for by extracting count rates from an
annular region around the central, piled-up location \footnote{Given
the pile-up and uncertainty in the XRT astrometry on the order of a
few arcseconds, attempting to separate any supernova flux from the other flux
is practically impossible}.

\section{Results}

\subsection{VLBI images at 22 GHz}
We imaged the data of SN\,2008iz from 2008 May 03 with a slightly 
super-resolved beam of 0.2 $\times$ 0.2 mas. The peak flux density was 5.9 
mJy~beam$^{-1}$, the total emission was 32 mJy and we achieved an image rms of 
79 $\mu$Jy~beam$^{-1}$. The data from 2009 April 08 were imaged with natural 
weighting and a circular beam of 0.5 $\times$ 0.5 mas to have more sensitivity 
for extended emission. Here, the peak flux density was 0.36 mJy~beam$^{-1}$, 
the total emission was 4.3 mJy, and we achieved an image rms of 
41 $\mu$Jy~beam$^{-1}$.

The supernova was clearly detected in both epochs (see 
Fig.~\ref{fig:hsa}). The source is already clearly resolved in the observation
on 2008 May 03 and shows a ring like structure, typical for a radio
supernova. In the following eleven months the source expanded and faded
significantly.

\subsection{VLA radio spectrum}

\begin{table}
\caption{Flux densities of SN\,2008iz and two other supernova remnants in M82 
from  the VLA observation on 2009 April 27.}
\label{tab:vla}      
\centering           
\begin{tabular}{cccccc}   
\hline\hline               
Source & F$_\mathrm{1.4\,GHz}$ & F$_\mathrm{4.8\,GHz}$& F$_\mathrm{8.4\,GHz}$& F$_\mathrm{22\,GHz}$ & F$_\mathrm{43\,GHz}$\\ 
&  [mJy]&  [mJy]&  [mJy]&  [mJy]&  [mJy]\\
\hline     
\\                   
SN\,2008iz &55.3$\pm$6.0 & 29.5$\pm$4.5 & 13.7$\pm$2.5 & 6.2$\pm$0.7 & 2.5$\pm$0.4\\
44.01+596& --& 14.7$\pm$5.3 & 15.0$\pm$2.0 & 8.7$\pm$0.6 & 3.6$\pm$0.4\\

45.17+612& -- &8.5$\pm$1.6 & 5.8$\pm$0.7 & 3.0$\pm$0.9 & 0.8$\pm$0.3 \\

\hline                                   
\end{tabular}
\end{table}

M82 was imaged at all frequencies using only data from 
baselines larger than 30 k$\lambda$. This ensures that most of the extended 
emission in M82 is resolved out. Flux densities were extracted by fitting 
two-dimensional Gaussians to the images.

The measured flux densities of SN\,2008iz and two other supernova remnants that
could be easily separated from the diffuse background emission are 
listed in Table~\ref{tab:vla}. The errors are estimated by adding in quadrature
the formal error from the fit to the images, the difference between peak 
and integrated flux densities, and an 5$\%$ error allowing for an error in the overall 
flux density scale. Furthermore, we added an additional 5$\%$ error at 1.4, 
4.8, and 8.4 GHz since we have more confusion from the extended emission at 
these frequencies. Note that the flux density at 4.8 GHz is in good 
agreement with the flux densities reported by \cite{BeswickMuxlowPedlar2009}, 
obtained with MERLIN a few days later (28.5$\pm$ 2 mJy).

\begin{figure}
   \resizebox{0.5\textwidth}{!}
            {\includegraphics[angle=-90]{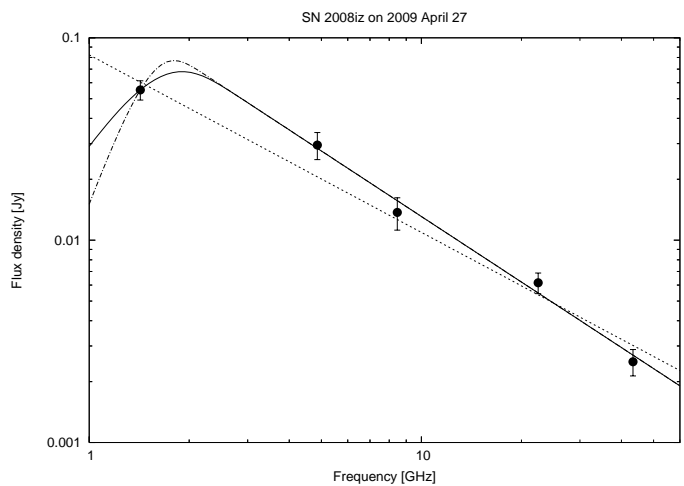}}
      \caption{Spectrum of SN\,2008iz taken with the VLA on 2009 April 27. Also
               shown are a single power-law fit (dotted line) and two broken 
               power-law fits with a spectral index of 2.5 (solid line) and 4.5
               (dash-dotted line) in the optically thick part.
   }
         \label{fig:vla}
   \end{figure}

The spectrum of SN\,2008iz is shown in Fig.~\ref{fig:vla}. First, we fitted a 
single power-law spectrum to the data. This gives a spectral index of 
$-0.88\pm$0.07. However, the fit has a large reduced $\chi^2$ value of 2.6, 
since the 1.4 GHz value is too low. Thus, we fitted the spectra also with a 
broken power-law,
\begin{equation}
S(\nu)=S_{0} \left(\frac{\nu}{\nu_0}\right)^\alpha\left(1-e^{-\left(\frac{\nu}{\nu_0}\right)^{\delta-\alpha}}\right),
\end{equation}
where $\alpha$ and $\delta$ are the spectral indices of the optically thin and
thick parts of the spectrum. $S_0$ and $\nu_0$ represent the maximum flux
density and the peak frequency of the fitted spectrum. Since one data point in
the optically thick part of the spectrum is not enough to fit the spectral
index there, we made two fits, one with a value for a synchrotron
self-absorbed spectrum ($\delta=$2.5), and a steeper free-free absorbed
spectrum ($\delta$=4.5). The reduced $\chi^2$ values in both cases are now
0.9. The fit (using $\delta=4.5$) gives a spectral index in the optically thin
part of $\alpha=-1.08\pm0.08$ and a turnover frequency of $\nu_0=1.51\pm0.09$
GHz. While the spectral index $\alpha$ is not affected by the choice of
$\delta$, $\nu_0$ changes slightly to 1.55 for $\delta=2.5$. This indicates
that the source was still optically thick, and brightening at the lowest
frequencies in April 2009 \cite[for comparision: SN\,1993J reached it's peak
at 1.4 GHz $\sim$500 days after the explosion;][]{Weiler2002}.

\subsection{X-ray upper limits}

The off-axis configuration and consequently the degraded angular resolution, 
as well as the diffuse background, have largely decreased the sensitivity of 
the Chandra measurements. The detection limit of these observations is
estimated  from the total emission around the radio position in a region with
the same  size as the point spread function. SN\,2008iz is located close to
several variable ultraluminous X-ray sources. However, no emission was
detected at the position of the supernova.  
\cite{MuxlowBeswickPedlar2009} and \cite{MuxlowBeswickGarrington2010}
report the discovery of a second radio transient in M82 with the MERLIN 
telescope. This source appeared between 2009 April 24 and 2009 May 5 and is 
located at a position with a diffuse emission background. It is 
surrounded by a few point-like sources. There is no enhanced X-ray emission 
at the location of this second radio transient, neither on 2009 April 17, nor 
on 2009 April 29.

To compare our limits with the emission from SN\,1993J, we assume spectral 
properties that are similar to the ones of SN\,1993J at a similar age, i.e., a
thermal bremsstrahlung spectrum with a  temperature of 1.05 keV, abundances
from  Table 2, Column 4 of \cite{ZimmermannAschenbach2003}, and an absorption
column density of 5.4$\times10^{22}$ cm$^{-2}$ (see Sect.~\ref{coldens}). Then
we get an 3 $\sigma$ upper limit of 1.5$\times 10^{39}$ erg~s$^{-1}$ in the
energy range 0.3-2.4 keV. This is consistent with the X-ray luminosity of
SN\,1993J at a similar age \citep{ZimmermannAschenbach2003} in the same energy
range.   

The 3 $\sigma$ sensitivity at the location of  the MERLIN transient, assuming
a column density of 10$^{22}$ cm$^{-2}$ and a photon power-law index of 1.7
\citep[an approximation to a thermal  bremsstrahlung spectrum with a
temperature of 10 keV, as seen in SN1995N;][]{FoxLewinFabian2000}, is found to
be about 1.2$\times$10$^{38}$ erg~s$^{-1}$ in 0.3--2.4 keV, which we take as
the upper limit of the X-ray luminosity.   

The resultant count rates in the Swift XRT data (0.5-8.0 keV)
are 0.60$\pm$0.02, 0.59$\pm$0.02, and 0.68$\pm$0.02 s$^{-1}$ on 2007 
January 26, 2008 May 1, and 2009 April 25, respectively.
There is no clear rise in total flux from the pre-SNe observation to
the May 2008 post-explosion epoch at $\sim$75 days, and while a $\sim$15\%
increase in count rate is seen in the final observation, this is
consistent with BeppoSAX observations of the central region of M82
showing variations on the order of 15-30\% (2-10 keV) on hour time-scales 
\citep{CappiPalumboPellegrini1999}. Indeed, such intrinsic variations are
also seen over the course of each epoch in the XRT observations as
well.   Given the complications with the data
and the intrinsic variability of the sources, strong constraints are
difficult.  However, since no increase in flux larger than the
intrinsic variability is seen, we can place an upper limit on the SNe
X-ray flux of approximately 1.5$\times10^{41}$ erg~s$^{-1}$ (0.5-2.4 keV) on 
2008 May 1 (assuming the same model as for the Chandra observations above).

\section{Expansion speed and explosion center}
\subsection{Estimates of size and position}
\label{size}

Estimating a physical size from the VLBI data is not straightforward.
Different methods have been applied to determine the expansion curves
of several radio supernovae \citep[e.g.][]{Bartel2002, Marcaide2009,
MarcaideMartiVidalPerezTorres2009}. Estimating the source size in the image 
plane may have {\em running biases} (i.e., biases that depend on the source 
size) related to the different resolutions achieved in the different images 
(i.e., the structure is better resolved as the source expands). One way to 
avoid such biases is to use a {\em dynamic beam}, i.e., a similar ratio between
source size and beam size \citep[e.g.][]{Marcaide1997} at each epoch. Here we 
use several different methods to determine the source size. The results for 
each method are summarized in Table~\ref{tab:methods}.

\begin{figure}
   \resizebox{0.5\textwidth}{!}
            {\includegraphics[angle=-90]{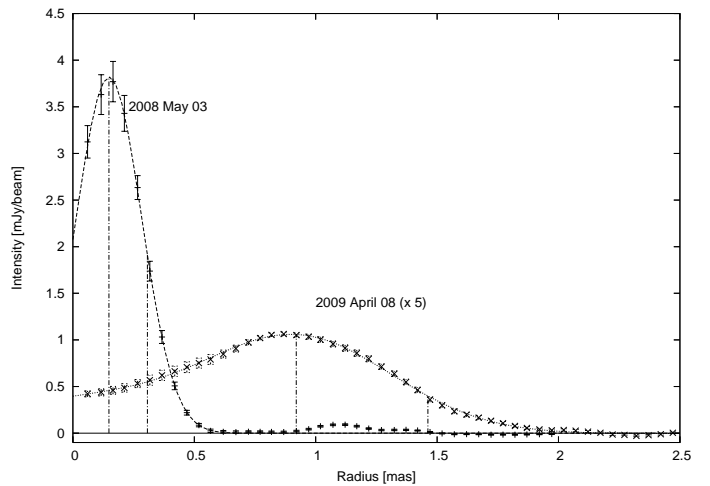}}
      \caption{Intensity of concentric rings of widths 0.05 mas in the images 
               from 2008 May 03 and 2009 April 08. The flux densities in the 
               second observation were scaled by a factor of 5. The vertical 
               lines mark the position of the peak and the radius where the 
               intensity is 50\% of the peak level.
   }
         \label{fig:rings}
   \end{figure}

\subsubsection{Concentric Rings}
\label{sizepos}

\begin{figure}
\centering
   \resizebox{0.55\textwidth}{!}
            {\includegraphics[bb=4.0cm 1.25cm 18.0cm 27.0cm,clip]{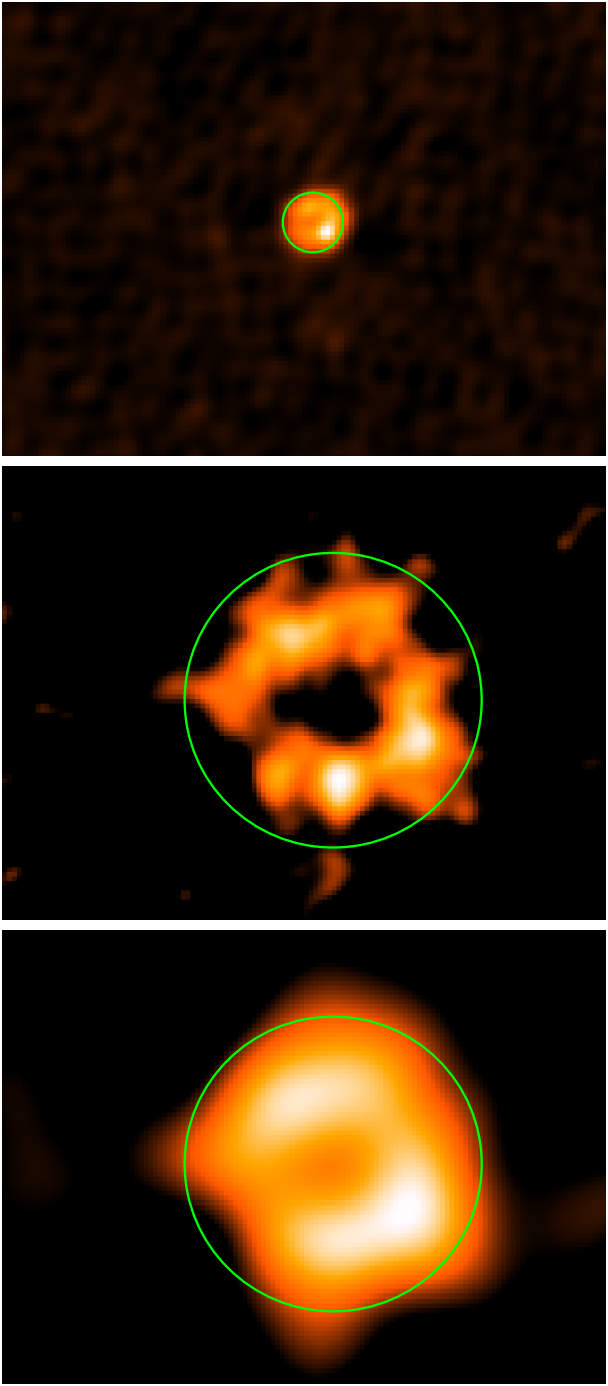}}

       \caption{HSA images at 22 GHz of SN\,2008iz in May 2008 (top) and 
        April 2009 (middle and bottom). The bottom image is convolved with a 
        dynamic beam of 0.99 mas (i.e. the ratios of beam size and source 
        size are identical in the top and bottom images). The rings denote
        the best fitted positions and radii from CPM  in both epochs, i.e. 320
        and 1580 $\mu$as.
   }
         \label{fig:hsa2}
   \end{figure}

First, we have measured the intensity in concentric rings (of 0.05 mas width) 
using the AIPS task IRING. Examples of intensity profiles for both observations
are shown in Fig.~\ref{fig:rings} together with fits to the data. One 
can define the radius R$_{\mathrm{n}\%}$ where the intensity falls to $n\%$ of 
the peak intensity, with an arbitrary value for $n$. Since R$_{\mathrm{n}\%}$ 
can be multi-valued (i.e., the inner and the outer radius), we always use the 
outer radius. Fig.~\ref{fig:rings} shows also the R$_{100\%}$ (i.e. the peak 
itself) and R$_{50\%}$ values. 
A value of $n$=50\% has the advantage that the profile is much steeper and the 
position can be better determined than at the peak (where the profile is flat).
In order to avoid a running bias, it is important to use in each epoch a 
similar ratio of source size to beam size (i.e. a dynamic beam). 
This method will give a size which is equal to the real size of the 
expanding shell times an unknown factor. This factor will be smaller for 
large values of $n$ (the measured sizes are smaller for large values of $n$).
Figure~\ref{fig:hsa2} shows the images of SN\,2008iz at both epochs and an 
image of the second epoch with a dynamic beam together with the two R$_{50\%}$ 
source sizes.

\begin{figure}
   \resizebox{0.5\textwidth}{!}
            {\includegraphics[angle=-90]{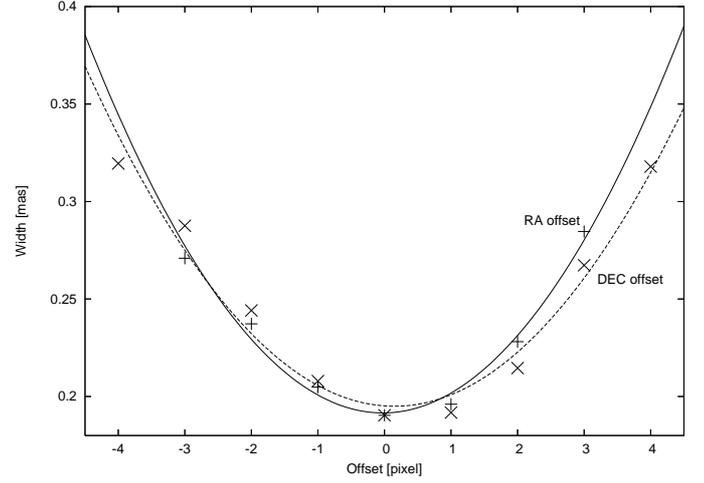}}
      \caption{The width of the intensity profile as a function of position 
               offset (relative to a reference pixel) in the observation on 
               2008 May 03 One pixel corresponds to 0.05 mas. Shown are 
               position offsets in right ascension (+) and declination (x). 
   }
         \label{fig:shift}
   \end{figure}

It is important to note that the intensity profile is sensitive to the position
of the central pixel of the concentric rings. A position offset would smear out
the intensity profile. This can be seen in Fig.~\ref{fig:shift} where we plot 
the width of the intensity profile for different position offsets in the first 
observation. As expected, the width shows a strong dependence on the position 
offset, and the position of the center of the shell can be estimated. The 
position of the center is located in the first epoch at 
09$^\mathrm{h}$55$^\mathrm{m}$51$^\mathrm{s}$.55026, 
+69$^\circ$40$'$45$''$.7913 (J2000). The uncertainty
in the absolute position is dominated by the uncertainty of the position of 
the phase referencing source M81* ($\sim$0.5 mas).

\subsubsection{Common Point Method}
\label{secCPM}

A different method that uses the concept of a dynamic beam in a natural way is 
the Common-Point Method \citep[CPM,][]{Marcaide2009}. This method relies on the
existence of a point in the radial profile of the supernova structure that 
remains unaltered under small changes in the convolving beam. This point is 
closely related to the source size \cite[indeed, the ratio between the radial 
position of this point and the source radius is $\sim1$; see Appendix A of][for
details]{Marcaide2009}. In Fig. \ref{CPM} we show the radial 
intensity profiles of SN\,2008iz, computed from the azimuthal averages of the 
CLEAN models convolved using different beams. The ``common points'' in the 
profiles can be clearly seen in the figure.

\begin{figure}
\centering
\includegraphics[width=9cm]{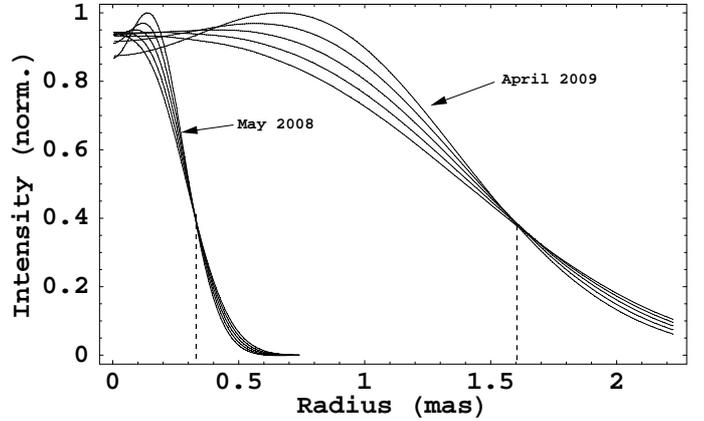}
\caption{Solid lines: radial intensity profiles of SN\,2008iz at the 
two VLBI epochs, computed using different convolving beams (1.0, 1.1, 1.2, 
1.3, and 1.4 times the source radius at each epoch). The radial positions 
of the common points at each epoch (see Section \ref{secCPM}) are marked with
dashed lines.} 
\label{CPM}
\end{figure}

\subsubsection{Model Fitting}

Finally, we estimated the size from model-fitting to the visibilities, 
using a simplified model of the supernova radio structure. This method may 
have undetermined running biases if the model fitted to the visibilities is 
not a good representation of the true supernova emission structure. 
We fitted the visibilities using a model of a spherical shell of 30\% 
fractional width. This is the model that best describes the 
structure of SN\,1993J in \cite{Marcaide2009}. A smaller fractional 
width between 20 and 25\%, as prefered by other authors 
\citep{BartelBietenholzRupen2000,Bartel2002,Bietenholz2003} or predicted by 
\cite{Chevalier1982}, could lead to different source sizes. However, the 
current data, based only on
two epochs at one frequency, do not allow a relaible statement about the
true fractional width of SN2008iz. For the model-fitting, we used 
a modified version of the subroutine {\sc modelfit} in the program {\sc difmap}
\citep{Shepherd1995}, which fits a parameterized shell model to the real and 
imaginary parts of the visibilities. The use of the real and imaginary parts 
of the visibilities (instead of amplitudes and phases) is more robust from a 
statistical point of view, especially in the case of a low signal-to-noise 
ratio, since the noise involved in the data is gaussian-like (as assumed 
in the modelling algorithms based on a $\chi^2$ minimization).

\subsection{Expansion speed}

Using the R$_{50\%}$ value as physical 
size, we get a radius of 307 $\mu$as on 2008 May 03 and 1462 $\mu$as on 2009 
April 08. The measured sizes correspond to an average expansion speed of 1243 
\nasyr\, or v$_\mathrm{exp}$=21200 \kms. Extrapolating this 
expansion back to a radius of zero and assuming constant expansion (r $\propto$
t$^m$ with an expansion index $m$=1) leads to an explosion date t$_0$ of 2008 
February 02. However, a significant deceleration (i.e. $m <$ 1) would shift the
explosion date to a later time. Since the supernova was first seen in a VLA 
observation 
on 2008 March 24, one can use this to give a lower limit for the expansion 
index $m$ of 0.7. \cite{MarchiliMartiVidalBrunthaler2009} model the 5 GHz 
light curve of SN\,2008iz and find evidence of a modest deceleration with an 
expansion index $m$=0.89  and an explosion date of 2008 February 18 ($\pm$6 
days). Extrapolating the expansion seen on the VLBI images backward using this 
expansion index, yields an explosion date of 2008 February 22. This value for 
the explosion date is in reasonable agreement with the estimate from the 5 GHz 
light curve.

The resulting source sizes for the different methods are summarized in 
Table~\ref{tab:methods}. Most methods give similar expansion speeds 
between 19600 and 23800 \kms. Since the R$_{100\%}$ values give a  
significantly lower expansion speed and a very late explosion date, we 
conclude that the R$_{100\%}$ values underestimate the true source sizes 
significantly. Hence, we do not consider these values for our further 
analysis.
Based on the information of the two VLBI images, we 
thus conclude that the explosion occured between 2008 January 22 and 2008 
March 24. We get a lower limit for the expansion index of $m>$0.65, and an 
average expansion speed in the range of 19600 -- 23800 \kms.

\begin{table}
\caption{Source sizes, expansion speeds, a lower limit for the expansion index 
$m$, and explosion dates (in 2008) assuming $m$=1 and $m$=0.89 for the different
methods outlined in Section \ref{size}.}
\label{tab:methods}      
\begin{tabular}{lcccccc}   
\hline\hline               
Method &R$_{2008}$&R$_{2009}$& v$_{\mathrm{exp}, m=1}$ &t$_0$ & $m$ & t$_0$\\
       & [$\mu$as] & [$\mu$as] & \kms & ($m$=1)& &($m$=0.89)\\
\hline
R$_{100\%}$   & 149 & 1008 & 15800 & Mar 05 & $>$0.86 & Mar 19\\
R$_{50\%}$    & 307 & 1462 & 21200 & Feb 02 & $>$0.70 & Feb 22 \\
R$_{25\%}$    & 380 & 1676 & 23800 & Jan 24 & $>$0.66 & Feb 14\\
CPM           & 320 & 1580 & 23100 & Feb 07 & $>$0.71 & Feb 25 \\
{\sc modelfit}& 320 & 1390 & 19600 & Jan 22 & $>$0.65 & Feb 12 \\
\hline

\hline                                   
\end{tabular}
\end{table}

Since the CPM source sizes should be very close to the real sizes, the 
expansion speed is very close to the average value of all 
methods ($\sim$22000 \kms, excluding R$_{100\%}$ which is clearly an outlier), 
and the explosion date is in very good agreement with the light curve
modeling, we adopt these values. The formal errors in our various fits are 
exceedingly small (of the order of 20 $\mu$as). Since the systematic errors
from the different methods are much larger, the spread in velocities
($\pm$1650 \kms) and explosion dates ($\pm$ 7 days) reflects our uncertainties.

This expansion speed is much higher than the values reported in 
\cite{BrunthalerMentenReid2009b}. This discrepancy has three reasons: i) The 
lower value was based on a preliminary data reduction, without the data 
from the phased VLA and without the geodetic block corrections. ii) The radius 
of the brightest emission was used as an approximation for the source size. 
However, as shown here, this method underestimates the true expansion speed. 
iii) The sizes were measured by hand from the images and this gave an 
overestimated size in the first epoch, where the source was still very compact.

\subsection{Expansion center, self-similarity, and anisotropic expansion}
\label{selfsim}

The ring is not symmetric and the brightest region in both 
images is the south-western part of the shell. This could be caused either by 
an asymmetry in the explosion or in the circumstellar medium (CSM), i.e. a 
clump with higher density in the southwest. A clumpy CSM is very common for red
supergiant stars \citep[e.g.][]{SmithHinkleRyde2009} which are likely 
progenitors for this supernova. Furthermore, a comparison of the 5 GHz 
light curve and the available 22 GHz data shows evidence for a clumpy CSM 
\citep{MarchiliMartiVidalBrunthaler2009}. 

\begin{figure}
\centering
\includegraphics[width=9cm]{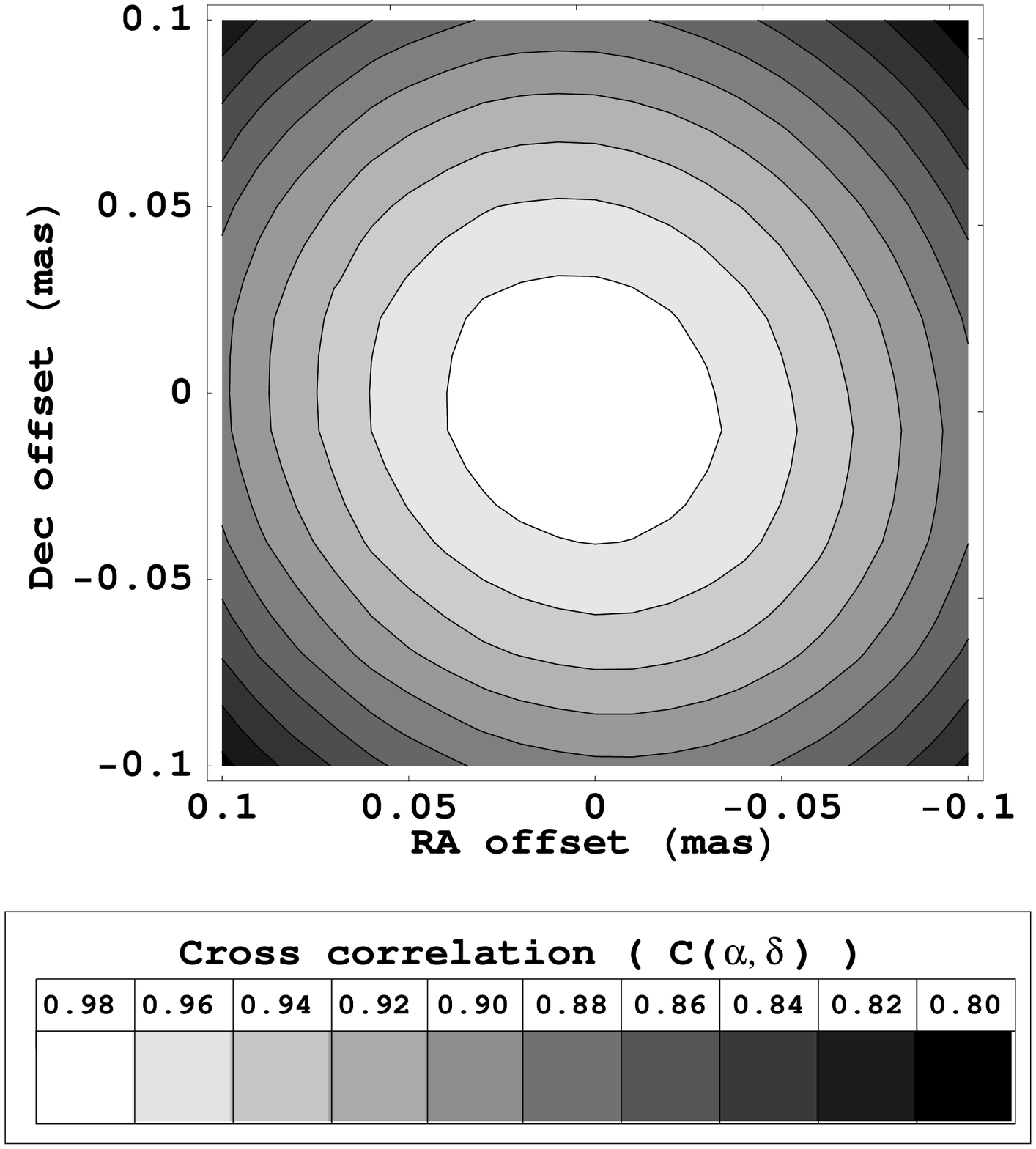}
\caption{Cross-correlation coefficient, $C(\alpha,\delta)$, between the images 
of the two VLBI epochs (scaling the first epoch to the size of the second), as 
a function of the coordinates of the explosion center, $(\alpha,\delta)$, 
taking as reference the position of the maximum $C(\alpha,\delta)$ (see Sect.~\ref{selfsim}).}
\label{CrossCorr}
\end{figure}

As mentioned in section~\ref{sizepos}, we also estimated the central position 
of the ring. We do not detect any significant shift in the position, with an
upper limit of $\sim50$ $\mu$as in right ascension and 100 $\mu$as in
declination. To verify the relative astrometric accuracy of the positions, we 
also imaged the three extragalactic background sources 0945+6924, 0948+6848, 
and 1004+6936. The average position change between the two observations of M81* 
relative to these quasars is 1$\pm$15 $\mu$as in right ascension and 22$\pm$53 
$\mu$as in declination, where the uncertainties are the standard 
deviation\footnote{These values give 1 $\sigma$ upper limits on the proper 
motion of M81 of 260 \kms\, in right ascension and 1270 \kms\, in declination}.
Thus, we can rule out that any significant position shift is introduced by jet 
motion in 
M81*. We note that the angular separation of M82 to M81* is not larger than 
the angular separation between M81* and the three background quasars. Hence, 
systematic errors in the astrometry should be similar for all sources.

According to \cite{Chevalier1982}, the expansion of a supernova
is self-similar, i.e., the source structure remains unaltered regardless of the
source size. This model of a self-similar expansion has been extensively tested
with SN\,1993J \citep[e.g.][]{Marcaide1997,Marcaide2009,Bartel2002}. 
Although some deviations from self-similarity were found in the expansion of 
SN\,1993J, this supernova kept its structure nearly self-similar during 
more than a decade. If the expansion of SN\,2008iz was also self-similar 
between our two VLBI epochs, the VLBI images obtained should be equal 
(regardless of the global flux density scale) if we scale the first image to 
the size of the second one and use the same convolving beam for both images. 
However, the result of expanding the first VLBI image, to compare to the 
second one, depends on the point in the image that is chosen as the center 
of expansion. Therefore, a direct comparison between the images, to check 
the self-similarity in the expansion, is not possible if the coordinates of 
the expansion center of the supernova are not known.

To determine the coordinates of the expansion center of the supernova 
independently from the method described above, we 
computed different expanded versions of the image of the first VLBI epoch 
using different centers of expansion. The scaled images were obtained by 
scaling the positions of the CLEAN components of the model, according to the 
ratios of the supernova sizes reported in the previous section. We then used 
the same beam to convolve the CLEAN model of the second epoch and the resulting
(scaled) CLEAN models of the first epoch. Finally, we compared the resulting 
image of the second epoch with the resulting (scaled) images of the first 
epoch by computing the cross correlation between both images using:
we mean

\begin{equation}
C(\alpha,\delta) = \frac{\sum_{i}^{}{I^i_2 \times I^i_1(\alpha,\delta)}}
{\sqrt{\sum_{i}^{}{\left(I^i_1(\alpha,\delta)\right)^2} \times \sum_{i}^{}{\left(I^i_2\right)^2}}}
\end{equation}

\noindent where $I^i_2$ is the $i$th pixel of the image of the second epoch
and $I^i_1(\alpha,\delta)$ is the $i$th pixel of the scaled 
version of the image of the first epoch, taking the point $(\alpha,\delta)$ 
as the expansion center. Assuming a self-similar expansion, the coordinates 
of the maximum value of $C(\alpha,\delta)$ are an estimate of the position of
expansion center of the supernova, which we identify as the coordinates 
of the explosion. In Fig. \ref{CrossCorr} we show the cross correlation of 
the images computed for different expansion centers, $(\alpha,\delta)$, 
expanding the image of the first epoch according to the size estimated with 
the CPM. The maximum correlation (which corresponds to our estimate of the 
explosion center) takes place at the coordinates 
$\alpha=09^\mathrm{h}\,55^\mathrm{m}\,51^\mathrm{s}.55025$ and 
$\delta=+69^{\circ}\,40'\,45''.79133$. 
These coordinates do not change by more than 10\,$\mu$as if we use, instead, 
the size estimated with model-fitting to expand the image of the first VLBI 
epoch.
This position agrees to the position given in Sect.~\ref{sizepos} within 60 
$\mu$as and verifies that the position of the expansion center has not changed
between the two epochs.

Additionally, the value of $C(\alpha,\delta)$ at the maximum is a measure of 
the degree of self-similarity (and/or anisotropy) in the expansion. The maximum
cross-correlation of the images is 0.98, if we use the CPM size to expand the
image of the first epoch, and 0.97, if we use instead the model-fitting size. 
These values are very close to 1, which is the case of a perfect self-similar 
expansion. Therefore, we conclude that the expansion of SN\,2008iz was 
self-similar to a high degree between our two VLBI epochs.

\section{Discussion}

\subsection{Column density and extinction}
\label{coldens}

\begin{figure}
\centering
\includegraphics[width=9cm]{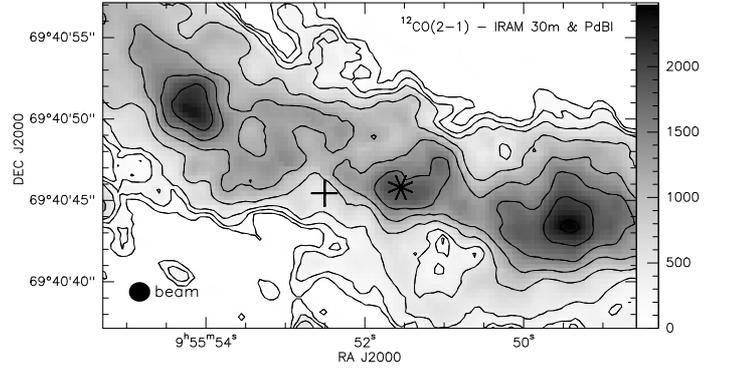}
\caption{$^{12}$CO (J=2$\rightarrow$1) intensity map of the central region of 
M82 from \cite{WeissNeiningerHuettenmeister2001} with the positions of 
SN\,2008iz (star) and the MERLIN transient (cross). The contours correspond to 
200, 400, 600, 800, 1200 1600, 2000, and 2400 K kms$^{-1}$. The resolution of
the observation is $1.5''\times 1.4''$ ($\sim$25 pc).}
\label{fig:co}
\end{figure}

The non-detection of SN\,2008iz in the optical, infrared, and X-rays indicates
that it exploded inside or behind a very dense cloud. Indeed, the $^{12}$CO 
(J=2$\rightarrow$1) line intensity map in 
\cite{WeissNeiningerHuettenmeister2001} 
shows a prominent cloud exactly at the position of the supernova
(Fig.~\ref{fig:co}). The line intensity at the position of SN\,2008iz is
$\sim$ 1800 K kms$^{-1}$. Using the Galactic conversion factor
X$_\mathrm{CO}=1.6\times 10^{20}$ cm$^{-2}$ (K kms$^{-1}$)$^{-1}$, this
corresponds to a H$_2$ column density $N(\rm{H}_2)$ of $\sim 29\times 10^{22}$
cm$^{-2}$. However, \cite{WeissNeiningerHuettenmeister2001} find much smaller
and spatially variable conversion factors from radiative transfer calculations, that lead to smaller
H$_2$ column densities. At the position of SN\,2008iz, the conversion factor
is $\sim  0.3\times10^{20}$ cm$^{-2}$ (K kms$^{-1}$)$^{-1}$ (their
Fig. 10). This leads to a H$_2$ column density of $\sim 5.4\times 10^{22}$
cm$^{-2}$. However, the CO observations were performed with a linear
resolution of $\sim$25 pc. Thus it is possible, that the supernova is located
behind a smaller cloud with much higher column density. 
The $^{12}$CO line intensity at the position of the
MERLIN transient is almost 4 times smaller ($\sim$500 K kms$^{-1}$).

Taking the latter value ($5.4\times 10^{22}$ cm$^{-2}$) for N$({\rm H}_2)$ and 
assuming that all of the column is between SN 2008iz and us, we derive a
visual extinction, $A_V$, of 24.4 mag. Here we have used the relation between
optical extinction and hydrogen nucleus column density, N(${\rm H}$), derived
by \citet{GueverOezel2009} from X-ray absorption data of a sample of Galactic
SNRs and assumed that all the hydrogen is in molecular form, i.e., N(${\rm
  H}_2$) = 2~N(${\rm H}$). Such a high value of the extinction would explain
the lack of an optical counterpart. 

The derived  extinction is much higher than toward SN\,1993J, for which
\cite{Richmond1994} discuss values of $A_V = 0.7$ and 1.0 mag. With the
extinction law given by \citet{Cardelli1989}, $A_K = 0.114~A_V$, we calculate
2.8 mag for the $K-$band ($2.2~\mu$m) extinction $A_K$.  
\cite{FraserSmarttCrockett2009} report an even higher $K$-band extinction of 
up to 11 mag based on their non-detection (3$\sigma$ upper limit on the \
absolute $K$-band magnitude of -5 mag) and the assumption that the infrared
light curve of SN\,2008iz behaves similar to the one of SN\,1993J. Thus, 
SN\,2008iz is either very weak in the infrared compared to SN\,1993J, or behind
a smaller but even denser cloud than estimated here.

\subsection{Expansion velocity and synchrotron self-absorption}

\cite{Chevalier1998} proposed a way to estimate the mean 
expansion velocity of a supernova based on the radio-light curve and 
assuming that synchrotron self-absorption (SSA) dominates the light 
curve at the peak. If SSA is not the dominant absorption process at 
the peak of the light curve, then the estimate of the expansion velocity
from the \cite{Chevalier1998} model is a lower bound to the real expansion 
velocity of the radio shell.  Following \cite{Chevalier1998}, the estimated 
mean expansion velocity at the peak of the radio-light curve, assuming 
dominant SSA, is:

\begin{eqnarray}
V_\mathrm{SSA}(\mathrm{km}\,\mathrm{s}^{-1}) = 5.3786\times10^{6}
({\beta\phi})^{-0.053}
\left (\frac{F_{\mathrm{p}}}{1 \mathrm{Jy}}\right )^{0.47}\times
\nonumber \\
\left (\frac{D}{1 \mathrm{ Mpc}}\right )^{0.95} \left 
(\frac{\nu}{1.0~\mathrm{ GHz}}\right )^{-1}
\left (\frac{t_{p}}{1 \mathrm{ day}}\right)^{-1}
\label{VelSSA}
\end{eqnarray}

\noindent where $\beta$ is the ratio of the relativistic particle energy 
density to the magnetic field energy density \cite[if we assume energy 
equipartition, $\beta = \frac{4}{3(1+k)}$, where $k$ ranges from 1 to 
2000, see ][chapter 7]{Pacholczyk1970}, $\phi$ is the filling factor 
of the emitting region to a sphere (we assume a shell of width equal to 
30\% of the outer radius, which yields $\phi = 0.66$), $F_{\mathrm{p}}$ is 
the flux density at the peak, $D$ is the distance, $\nu$ is the observing 
frequency, $t_{p}$ is the supernova age at the peak, and we use a spectral 
index $\alpha = -1$ \cite[see Eqs. 11 and 13 of ][]{Chevalier1998}.

For SN2008iz, Eq. \ref{VelSSA} yields a mean expansion velocity (depending 
on k) in the range $8100 - 11800$\,km\,s$^{-1}$ at the peak 
of the 5\,GHz radio light curve. Using an expansion index of 0.89, 
these velocities translate into a mean expansion velocity in the 
range $7000 - 10300$\,km\,s$^{-1}$ at the epoch of 27 April 2009. 
This range of velocities is a factor $\sim 2$ smaller than the
velocities estimated from our VLBI observations, thus indicating 
that, in contrast to the case of SN1993J, SSA may not be 
an important absorption mechanism in the SN2008iz radio emission.
This is also consistent with the results in 
\cite{MarchiliMartiVidalBrunthaler2009} who were able to model the radio light 
curve of SN2008iz assuming that SSA effects are much smaller than free-free 
absorption (FFA) during the whole supernova expansion.

\subsection{Comparison with other type II radio supernovae}

\begin{table}
\caption{Comparison between SN\,1993J and SN\,2008iz.}
\label{tab:comp}      
\begin{tabular}{lccc}   
\hline\hline               
Property&&SN\,1993J&SN\,2008iz \\
\hline
spectral index $\alpha$& &-0.99$^a$&-1.08$^b$\\
L$_{5 \mathrm{GHz}}$& $[10^{27}$erg~s$^{-1}$~Hz$^{-1}]$&1.5$^c$&2.5$^d$\\

t$_\mathrm{peak}$-t$_0$& [days]&180$^c$&$\sim$120$^d$\\
v$_\mathrm{VLBI}$&[\kms]&14900$^e$ & 21200$^b$\\
L$_\mathrm{X-ray\,at\,t\sim220\,days}$& $[10^{38}$ erg~s$^{-1}]$&$\sim8^f$ &$<15^b$\\
\hline                                   
\end{tabular}

References: a) \cite{vanDykWeilerSramek1994}; b) this work; c) \cite{WeilervanDykMontes1998}; d) \cite{MarchiliMartiVidalBrunthaler2009};  e) \cite{MarcaideAlberdiRos1995}; f) \cite{ZimmermannAschenbach2003}.
\end{table}

Estimates of the expansion velocities of other type II radio supernovae
have been estimated from VLBI observations, and very different results
have been obtained. For instance, the mean expansion velocity of 
SN\,1979C during the first year after explosion is estimated to be $\sim
10000 - 11000$ \kms\, \citep{BartelBietenholz2003,
MarcaideMartiVidalPerezTorres2009}; for 
SN\,1986J, a velocity of $\sim$ 14700 \kms\, was obtained 
by \cite{PerezTorresAlberdiMarcaide2002}, while 
\cite{BietenholzBartelRupen2002} find 20000 \kms\, 3 month 
after the explosion; \cite{Staveley-SmithBriggsRowe1993} report a mean 
expansion speed of $\sim$ 35000 \kms\ during the first years for SN\,1987A
before it slowed down to $\sim$ 4800 \kms; 
for SN\,2004et, the expansion velocity was $>$ 15700 \kms\, 
\citep{MartiVidalMarcaideAlberdi2007}; and for SN\,2008ax, an
expansion velocity as large as 52000 \kms\, was obtained
\citep{MartiVidalMarcaideAlberdi2009}. Estimates of the 
expansion velocities of other supernova remnants in M82 (the 
host galaxy of SN\,2008iz), have been also reported, which range 
between $\sim$ 1500 and 11000 \kms\, \citep{BeswickRileyMartiVidal2006}. 
These later velocities are much higher (a 
factor of $3-22$) than the predicted velocities from the model of
\cite{ChevalierFransson2001}, based on the high 
pressure expected in the interstellar medium (ISM) of M\,82. The
expansion velocity reported in this paper for SN2008iz is indeed a
factor $\sim 40$ larger than the predicted velocities in 
\cite{ChevalierFransson2001}, although of the same 
order of magnitude than the velocities reported in 
\cite{BeswickRileyMartiVidal2006} for the other remnants in M82, and the 
typical velocities of the other type II supernovae observed to date.

\cite{WeilervanDykMontes1998} find a correlation between peak radio luminosity
at 5 GHz and  the time between the explosion and the peak in the 5 GHz 
light curve for type II supernovae. The 5 GHz light curve of SN\,2008iz from 
\cite{MarchiliMartiVidalBrunthaler2009} gives a peak luminosity of 
$\sim 2.5\times10^{27}$erg~s$^{-1}$~Hz$^{-1}$ at $\sim$120 days after the 
explosion. These values are well within the scatter of the correlation. Thus 
it seems plausible that SN\,2008iz is also a type II supernova.

Since SN\,2008iz and SN\,1993J  are located at very similar distances, this 
allows a 
detailed comparison between these two supernovae. Several properties of 
both supernovae are summarized in Table~\ref{tab:comp}. The radio 
spectral indices, the peak radio luminosities, rise times, and early VLBI 
expansion velocities are similar (considering that the rise times and peak 
radio luminosities can vary by several orders of magnitudes for type II radio 
supernovae). The non-detection in X-rays of SN\,2008iz can be attributed to 
absorption by the dense molecular cloud seen in the CO data.

\section{Summary}
In this paper we presented the first VLBI images, a VLA radio spectrum from 
1.4 to 43 GHz, and Chandra X-ray observations of SN\,2008iz. Our main results
are:
\begin{itemize}
\item The VLBI images, separated by $\sim$ 11 month show a shell-like structure
expanding with a velocity of $\sim$ 23000 \kms.
\item The inferred expansion speed is a factor of 2 higher than expected if SSA 
dominates the light curve. This indicates that SSA is not important for the 
radio emission.
\item The most likely explosion date is in mid February 2008, but not earlier 
than January 22 and not later than March 24.
\item We find no evidence for an asymmetric explosion, but a high degree of 
self-similarity between the two VLBI observations.
\item The VLA radio spectrum is well fitted by a broken power law with a
turnover frequency of 1.5$\pm$0.1 GHz, and a spectral index of $-1.08\pm0.08$
in the optically thin part. 
\item SN\,2008iz is located behind (or inside) a large molecular cloud with a
H$_2$ column density of 5.4$\times10^{22}$ cm$^{-2}$ (on a scale of 25 pc),
corresponding to a visual extinction, $A_V$, of 24.4 mag.
\item Due to the high column density, we obtain only upper limits on the
X-ray luminosity of 1.5$\times10^{41}$  and 1.5$\times10^{39}$erg~s$^{-1}$ 
$\sim$75 and $\sim$200 days after the explosion, which is consistent with 
the X-ray luminosity of SN\,1993J at similar ages. We also obtain an 
upper limit for the X-ray luminosity of the second radio transient of 
1.2$\times$10$^{38}$ erg~s$^{-1}$ (in 0.3--2.4 keV).
\end{itemize}

\begin{acknowledgements}
This work is partially based on observations with the 100-m telescope of the 
MPIfR (Max-Planck-Institut f\"ur Radioastronomie) at Effelsberg. MV is a fellow
of the Alexander von Humboldt Foundation in Germany.
\end{acknowledgements}

\bibliographystyle{aa}
\bibliography{brunthal_refs}

\end{document}